\documentclass[%
 reprint,
nofootinbib,
amsmath,amssymb,
aps,
pra,
floatfix,
]{revtex4-2}

\usepackage{amsmath}
\usepackage{amssymb}
\usepackage{xcolor}
\usepackage{color}
\usepackage{titlesec}
\usepackage{graphicx}
\usepackage[export]{adjustbox}
\usepackage{ragged2e} 

\usepackage{flushend}
\usepackage{booktabs}
\usepackage{tabularx}
\usepackage{multirow}
\usepackage{dcolumn}
\usepackage{bm}
\usepackage{algorithm2e}
\usepackage{algorithmic}%
\usepackage{ulem}
\RestyleAlgo{ruled}
\usepackage{appendix}

\usepackage{hyperref}

\usepackage{silence}
\begin{document}
\setlength\abovedisplayskip{6pt}
\setlength\belowdisplayskip{6pt}
\setlength\abovedisplayshortskip{4pt}
\setlength\belowdisplayshortskip{4pt}


\title{Quantum Noise Suppression Beyond the Standard Quantum Limit in a Hybrid Magnonic
Optomechanical System}
\author{Alolika Roy , Amarendra K. Sarma }
{ \affiliation{ 
     Department of Physics, Indian Institute of Technology Guwahati, Guwahati 781039, India.}
\date{\today}
\begin{abstract}

We theoretically study how quantum measurement noise can be engineered in a hybrid cavitymagnomechanical platform for precision force sensing. The proposed configuration consists of a driven optomechanical cavity, with a movable mirror on one side plus a fixed semi-transparent mirror on the other side, coupled to a magnon mode, with an OPA placed inside the cavity. We show that the magnon mediated dynamics reshapes the added-noise spectrum leading to improved sensitivity compared to a conventional optomechanical sensor. In particular, by satisfying the coherent quantum noise cancellation (CQNC) criterion, radiation-pressure back-action can be fully suppressed. In addition, a larger OPA pump gain permits operation beyond the standard quantum limit at substantially reduced laser power, thereby mitigating power-related constraints without sacrificing performance. These combined advantages provide a practical pathway to below-SQL weak force detection and can outperform existing approaches based on squeezing in magnomechanics.

\end{abstract}
%


\maketitle

\section{Introduction}

Quantum optomechanics has emerged as a powerful platform for precision sensing and metrology, with important applications in the detection of extremely weak forces and displacements \citep{caves1980measurement,danilishin2012quantum,RevModPhys.86.1391,LiOuLeiLiu+2021+2799+2832}. A variety of quantum-optical interference and correlation effects, including quantum correlations, coherent population trapping, electromagnetically induced transparency, and noise squeezing, have played an important role in tailoring the balance between signal and noise in cavity-based measurements \citep{bemani2019quantum,Gray:78,PhysRevLett.66.2593,Liu2001,PhysRevD.23.1693,RevModPhys.77.633,Zhu:20,dalafi2018effects}. These developments have established optomechanical systems as versatile platforms for exploring the limits of quantum-enhanced sensing \citep{PhysRevLett.96.010401,chen2013macroscopic}.

In standard cavity optomechanics, however, measurement sensitivity is fundamentally limited by the competition between shot noise and radiation-pressure back-action noise. Since these two contributions scale oppositely with the driving power, their interplay gives rise to the standard quantum limit (SQL), which remains a central benchmark in precision measurement \citep{bowen2015quantum,caves1980measurement,clerk2010introduction,danilishin2012quantum}. Reaching or surpassing this limit has been a central objective in quantum sensing for several decades \citep{braginsky1980vorontsov}. In particular, although increasing the laser power reduces shot noise, it simultaneously enhances radiation-pressure fluctuations, making back-action suppression essential for further improvement in force sensitivity \citep{Carney_2021,schreppler2014optically,mason2019continuous}.

Several strategies have been developed to overcome this limitation, including coherent quantum noise cancellation (CQNC) \citep{tsang2010coherent}, back-action-evading measurement schemes \citep{ghosh2019back,Shomroni2019,bittencourt2025magnon}, squeezed-light-enhanced detection \citep{PhysRevLett.59.278,Hoff:13,chelkowski2005experimental}, inverted atomic-spin systems \citep{hammerer2009establishing}, and negative-mass oscillator approaches \citep{zhang2013back,moller2017quantum,PhysRevLett.121.031101}. These approaches rely on engineering additional interference pathways or measurement protocols that suppress radiation-pressure back-action while preserving the signal of interest \citep{murch2008observation,purdy2013observation,kimble2001conversion,chen2011qnd}. Among them, CQNC is particularly appealing because it suppresses back-action dynamically through destructive interference with an auxiliary mode possessing an appropriately matched response \citep{bondurant1986reduction,tsang2010coherent,wimmer2014coherent}.

Hybrid optomechanical systems are especially well suited for implementing such noise-cancellation strategies, since additional degrees of freedom can be introduced and engineered to provide the required interference channels \citep{barzanjeh2011entangling,stannigel2010optomechanical,stannigel2011optomechanical,stannigel2012optomechanical,xiang2013hybrid}. In this context, magnon-assisted hybrid systems are particularly appealing. Magnon excitations are bosonic collective spin modes and can serve as coherent ancillary degrees of freedom in hybrid quantum platforms \citep{PhysRev.58.1098,Gupta:24}. Their spectral selectivity and tunability make them promising candidates for interference-based control of measurement back-action. Recent progress in cavity optomagnonics has demonstrated photon--magnon coupling in whispering-gallery-mode and integrated photonic platforms, while related studies have explored microdisk-based geometries and mode engineering for enhancing optomagnonic interactions \citep{Zhu2020WaveguideOptomagnonics,Chai2022WGMMicrocavityConversion,Wu2021PhotonMagnonMicrodisk,Wang2022OptomagnonicsReview}. These systems offer tunability through magnetic bias, mode selection, and device design, which makes magnons attractive ancillary modes for interference-based noise engineering. These features motivate the exploration of magnon-assisted CQNC in force-sensing architectures.

Motivated by these developments, in this work we study a hybrid cavity magnomechanical architecture in which a single optical cavity is coupled simultaneously to a mechanical oscillator through radiation pressure and to a magnon mode, while an intracavity optical parametric amplifier (OPA) is used to further tailor the measurement noise. We show that, by appropriately tuning the cavity detuning, the effective linearized couplings, and the magnon linewidth, the system can realize coherent quantum noise cancellation \citep{tsang2010coherent,bondurant1986reduction}. In this regime, the magnon-mediated pathway provides an equal-and-opposite contribution to the measurement back-action, thereby suppressing the added force noise and enabling below-SQL sensitivity near the mechanical resonance \citep{Motazedifard_2016,bariani2015atom,singh2023enhanced,allahverdi2022homodyne,wimmer2014coherent}.

Using a linearized quantum-Langevin treatment and frequency-domain response analysis, we derive analytical expressions for the output field quadratures and the corresponding added-force noise spectrum. We identify the matching conditions required for ideal CQNC and examine the effect of realistic imperfections in coupling strengths and dissipation rates. In addition, we analyze the role of the intracavity OPA and show that phase-sensitive gain can reduce the imprecision background and shift the optimal sensing regime toward lower input laser power \citep{peano2015intracavity,Motazedifard_2016,singh2023enhanced}. Such low-power operation is experimentally attractive because it helps mitigate practical limitations associated with heating, instability, and technical noise in precision optomechanical measurements \citep{PhysRevLett.94.223902,roy2025overcoming}.

Overall, our results show that the combined action of a magnon ancilla and intracavity parametric amplification provides a promising route toward back-action-suppressed, low-power force sensing in hybrid quantum platforms. Compared with standard optomechanical sensing schemes, the proposed cavity magnomechanical CQNC setup achieves substantially improved sensitivity and offers a complementary route to previously studied weak-force sensing strategies based on squeezed magnomechanics \citep{zhang2024quantum}.

\section{System}
\label{system}

\begin{figure}[h!]
\centering
\setlength{\fboxsep}{0pt}
\setlength{\fboxrule}{0.6pt}
\fbox{\includegraphics[height=5cm,clip]{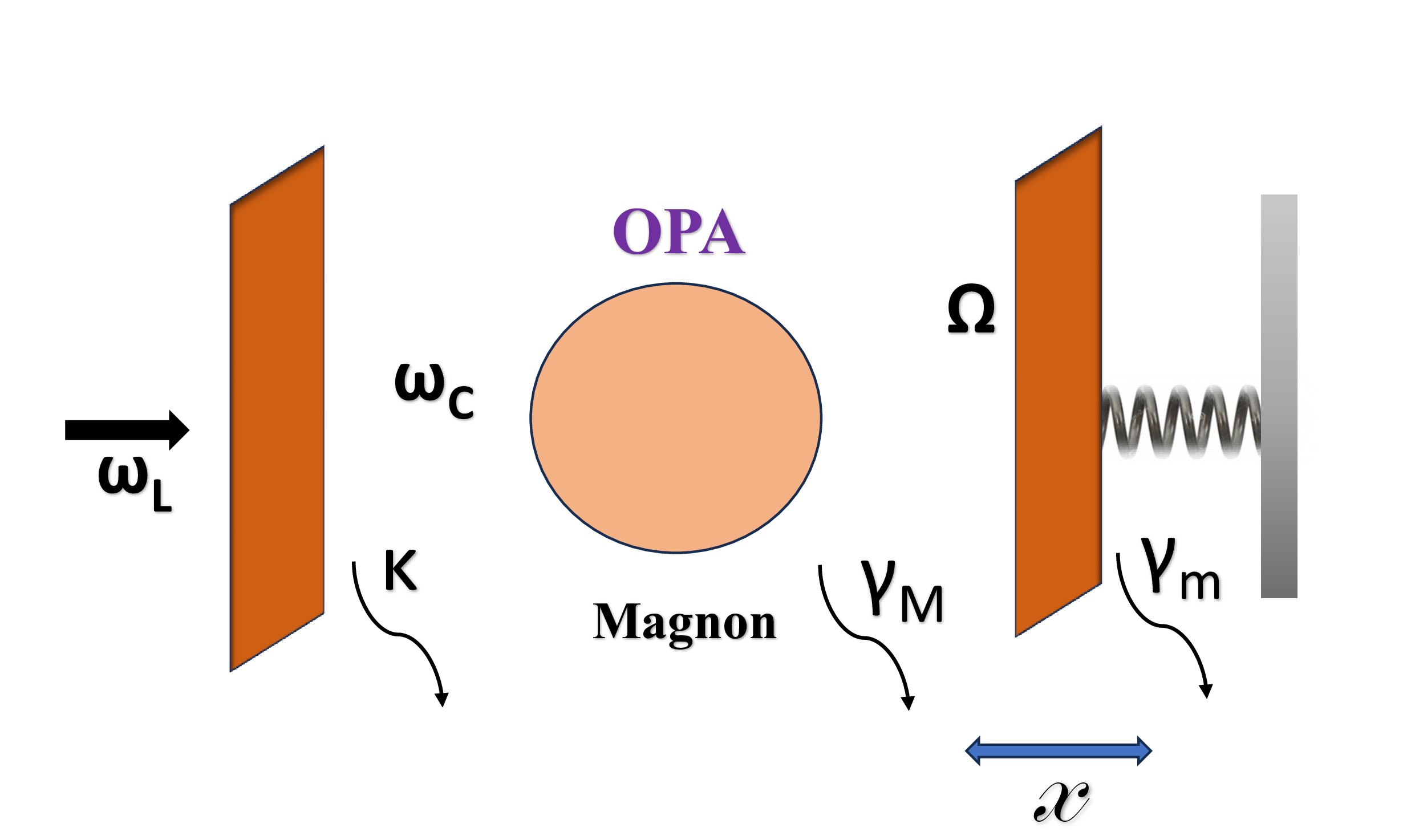}}
\caption{Schematic diagram of the hybrid cavity magnomechanical platform equipped with a magnon mode and an intracavity optical parametric amplifier (OPA).}
\label{fig:schematic_omm}
\end{figure}

We consider a hybrid cavity magnomechanical system consisting of a single optical cavity mode of resonance frequency \(\omega_c\), a mechanical oscillator of frequency \(\Omega\), and a magnon mode of frequency \(\omega_M\). The cavity field couples to the mechanical oscillator through the standard radiation-pressure interaction and to the magnon mode through a linearized cavity-magnon interaction. In addition, the cavity contains an intracavity optical parametric amplifier (OPA) and is driven coherently by an external laser of frequency \(\omega_L\) and input power \(P_L\). A schematic of the setup is shown in Fig.~\ref{fig:schematic_omm}.

The total Hamiltonian of the system can be written as
\begin{equation}
H = H_{\mathrm{OM}} + H_{\mathrm{OPA}} + H_{\mathrm{L}} + H_{\mathrm{M}} .
\label{eq:Htotal}
\end{equation}

The optomechanical part is
\begin{equation}
H_{\mathrm{OM}}
=
\hbar \Omega \hat{b}^{\dagger}\hat{b}
+
\hbar \omega_c \hat{a}^{\dagger}\hat{a}
-
\hbar g_0 \hat{a}^{\dagger}\hat{a}\,(\hat{b}+\hat{b}^{\dagger}),
\label{eq:Hom}
\end{equation}
where \(\hat a\) (\(\hat a^\dagger\)) and \(\hat b\) (\(\hat b^\dagger\)) are the annihilation (creation) operators of the cavity and mechanical modes, respectively, and \(g_0\) denotes the single-photon optomechanical coupling strength.

The OPA contribution is
\begin{equation}
H_{\mathrm{OPA}}
=
i\hbar \mathcal{G}
\left(
\hat{a}^{\dagger 2} e^{i\theta}
-
\hat{a}^{2} e^{-i\theta}
\right),
\label{eq:Hopa}
\end{equation}
where \(\mathcal{G}\) is the parametric gain and \(\theta\) is the pump phase. For simplicity, we take \(\theta=0\), so that
\begin{equation}
H_{\mathrm{OPA}}
=
i\hbar \mathcal{G}
\left(
\hat{a}^{\dagger 2}
-
\hat{a}^{2}
\right).
\label{eq:Hopa_theta0}
\end{equation}

The coherent laser drive is described by
\begin{equation}
H_{\mathrm{L}}
=
i\hbar E_L (\hat a^\dagger-\hat a),
\label{eq:Hdrive}
\end{equation}
where \(E_L\) is the amplitude of the driving field.

The magnonic part of the Hamiltonian is written as
\begin{equation}
H_{\mathrm{M}}
=
\hbar \omega_M \hat M^\dagger \hat M
+
\frac{\hbar G_{OM}}{2}
(\hat a+\hat a^\dagger)(\hat M+\hat M^\dagger),
\label{eq:Hm}
\end{equation}
where \(\hat M\) (\(\hat M^\dagger\)) denotes the annihilation (creation) operator of the magnon mode, and \(G_{OM}\) is the cavity-magnon coupling strength.

Moving to the frame rotating at the drive frequency \(\omega_L\), and dropping constant terms, the total Hamiltonian becomes
\begin{equation}
\begin{aligned}
H
&=
-\hbar \Delta_c \hat a^\dagger \hat a
+
\hbar \Omega \hat b^\dagger \hat b
-
\hbar \Delta_M \hat M^\dagger \hat M
+
i\hbar \mathcal{G}
\left(
\hat a^{\dagger 2} e^{i\theta}
-
\hat a^2 e^{-i\theta}
\right)
\\[2mm]
&\quad
-
\hbar g_0 \hat a^\dagger \hat a (\hat b+\hat b^\dagger)
+
\frac{\hbar G_{OM}}{2}
(\hat a+\hat a^\dagger)(\hat M+\hat M^\dagger)\\
&\quad
+
i\hbar E_L(\hat a^\dagger-\hat a),
\end{aligned}
\label{eq:Hrot}
\end{equation}
where
\begin{equation}
\Delta_c=\omega_L-\omega_c,
\qquad
\Delta_M=\omega_L-\omega_M.
\end{equation}

We now linearize the dynamics around the semiclassical steady state by writing each system operator as
\begin{equation}
\hat A(t)=\bar A+\delta \hat A(t),
\end{equation}
where \(\bar A\) denotes the steady-state mean value and \(\delta \hat A(t)\) represents the corresponding quantum fluctuation. Retaining only terms up to first order in the fluctuations yields the linearized description of the system.

For each bosonic mode \(A=a,b,M\), we define the amplitude and phase quadratures as
\begin{equation}
\hat x_A=\frac{\hat A+\hat A^\dagger}{\sqrt{2}},
\qquad
\hat p_A=i\frac{\hat A^\dagger-\hat A}{\sqrt{2}}.
\label{eq:quadratures}
\end{equation}
Similarly, for the input noise operators \(\hat A_{\mathrm{in}}\), we define
\begin{equation}
\hat x_{A,\mathrm{in}}
=
\frac{\hat A_{\mathrm{in}}+\hat A_{\mathrm{in}}^\dagger}{\sqrt{2}},
\qquad
\hat p_{A,\mathrm{in}}
=
i\frac{\hat A_{\mathrm{in}}^\dagger-\hat A_{\mathrm{in}}}{\sqrt{2}}.
\label{eq:inputquadratures}
\end{equation}
The input noises satisfy the Markovian correlations \citep{walls1994gj,benguria1981quantum,gardiner1985input,wimmer2014coherent}
\begin{equation}
\langle \hat a_{\mathrm{in}}(t)\hat a_{\mathrm{in}}^\dagger(t')\rangle
=
\langle \hat b_{\mathrm{in}}(t)\hat b_{\mathrm{in}}^\dagger(t')\rangle
=
\langle \hat M_{\mathrm{in}}(t)\hat M_{\mathrm{in}}^\dagger(t')\rangle
=
\delta(t-t').
\label{eq:noisecorr}
\end{equation}

The resulting linearized quantum Langevin equations are
\begin{align}
\dot{\hat x}_b
&=
\Omega \hat p_b,
\label{eq:qle1}
\\
\dot{\hat p}_b
&=
-\Omega \hat x_b
-\gamma_m \hat p_b
-g\hat x_a
+
\sqrt{2\gamma_m}\,
\bigl(\hat F_{\mathrm{th}}+F_{\mathrm{ext}}\bigr),
\label{eq:qle2}
\\
\dot{\hat x}_a
&=
\left(-\frac{\kappa}{2}+2\mathcal G\right)\hat x_a
+
\sqrt{\kappa}\,\hat x_{a,\mathrm{in}},
\label{eq:qle3}
\\
\dot{\hat p}_a
&=
-g\hat x_b
+\left(-\frac{\kappa}{2}-2\mathcal G\right)\hat p_a
-G_{OM}\hat x_M
+
\sqrt{\kappa}\,\hat p_{a,\mathrm{in}},
\label{eq:qle4}
\\
\dot{\hat x}_M
&=
-\frac{\gamma_M}{2}\hat x_M
-\Delta_M \hat p_M
+
\sqrt{\gamma_M}\,\hat x_{M,\mathrm{in}},
\label{eq:qle5}
\\
\dot{\hat p}_M
&=
-G_{OM}\hat x_a
+\Delta_M \hat x_M
-\frac{\gamma_M}{2}\hat p_M
+
\sqrt{\gamma_M}\,\hat p_{M,\mathrm{in}}.
\label{eq:qle6}
\end{align}
Here,
\begin{equation}
g=\sqrt{2}\,g_0\,\alpha,
\end{equation}
with \(\alpha=\langle \hat a\rangle\) denoting the steady-state intracavity field amplitude. The parameters \(\gamma_m\), \(\kappa\), and \(\gamma_M\) are the mechanical damping rate, cavity decay rate, and magnon damping rate, respectively. The terms \(\hat F_{\mathrm{th}}\) and \(F_{\mathrm{ext}}\) denote the Brownian thermal force and the external force acting on the mechanical oscillator.

To obtain the response functions, we move to the frequency domain using the Fourier transform
\begin{equation}
\hat O(\omega)
=
\frac{1}{\sqrt{2\pi}}
\int dt\,\hat O(t)e^{-i\omega t}.
\label{eq:FT}
\end{equation}
The linearized Langevin equations then take the form
\begin{align}
\hat X_b(\omega)
&=
\chi_m(\omega)
\left[
-g\,\hat X_a(\omega)
+
\sqrt{2\gamma_m}
\bigl(\hat F_{\mathrm{th}}(\omega)+F_{\mathrm{ext}}(\omega)\bigr)
\right],
\label{eq:freq1}
\\
\hat P_b(\omega)
&=
\frac{i\omega}{\Omega}\hat X_b(\omega),
\label{eq:freq2}
\\
\hat X_a(\omega)
&=
\sqrt{\kappa}\,\lambda_+(\omega)\hat X_{a,\mathrm{in}}(\omega),
\label{eq:freq3}
\end{align}

\begin{align}
\hat P_a(\omega)
&=
g^2\chi_m(\omega)\lambda_-(\omega)\lambda_+(\omega)\sqrt{\kappa}\,\hat X_{a,\mathrm{in}}(\omega)
\nonumber\\
&\quad
-g\chi_m(\omega)\lambda_-(\omega)\sqrt{2\gamma_m}
\bigl(\hat F_{\mathrm{th}}(\omega)+F_{\mathrm{ext}}(\omega)\bigr)
\nonumber\\
&\quad
-G_{OM}\lambda_-(\omega)\hat X_M(\omega)
+
\lambda_-(\omega)\sqrt{\kappa}\,\hat P_{a,\mathrm{in}}(\omega),
\label{eq:freq4}
\\
\hat X_M(\omega)
&=
-\Delta_M\chi_M(\omega)\hat P_M(\omega)
+
\chi_M(\omega)\sqrt{\gamma_M}\,\hat X_{M,\mathrm{in}}(\omega),
\label{eq:freq5}
\\
\hat P_M(\omega)
&=
-G_{OM}\,\xi_M(\omega)\lambda_+(\omega)\sqrt{\kappa}\,\hat X_{a,\mathrm{in}}(\omega)
\nonumber\\
&\quad
-\chi_M'(\omega)\sqrt{\gamma_M}\,\hat X_{M,\mathrm{in}}(\omega)
+
\xi_M(\omega)\sqrt{\gamma_M}\,\hat P_{M,\mathrm{in}}(\omega).
\label{eq:freq6}
\end{align}
Here we have introduced the susceptibilities and auxiliary response functions
\begin{align}
\chi_m(\omega)
&=
\frac{\Omega}{\Omega^2-\omega^2+i\gamma_m\omega},
\\
\chi_a(\omega)
&=
\frac{1}{i\omega+\kappa/2},
\\
\chi_M(\omega)
&=
\frac{1}{i\omega+\gamma_M/2},
\\
\xi_M(\omega)
&=
\left(
i\omega+\frac{\gamma_M}{2}+\Delta_M^2\chi_M(\omega)
\right)^{-1},
\\
\chi_M'(\omega)
&=
-\Delta_M \xi_M(\omega)\chi_M(\omega),
\\
\lambda_\pm(\omega)
&=
\left(\chi_a^{-1}(\omega)\mp 2\mathcal G\right)^{-1}.
\end{align}
For later convenience, we also define the effective linearized cavity-magnon coupling
\begin{equation}
G_{OM}'=\sqrt{2}\,G_{OM}\,\alpha_M,
\end{equation}
where \(\alpha_M=\langle \hat M\rangle\) denotes the steady-state magnon amplitude.

The dimensionless external and thermal forces are defined as
\begin{equation}
F_{\mathrm{ext}}
=
\frac{f_{\mathrm{ext}}}{\sqrt{\hbar m\Omega\gamma_m}},
\qquad
F_{\mathrm{th}}
=
\frac{f_{\mathrm{th}}}{\sqrt{\hbar m\Omega\gamma_m}},
\label{eq:scaledforces}
\end{equation}
where \(f_{\mathrm{ext}}\) and \(f_{\mathrm{th}}\) are the corresponding physical forces \citep{allahverdi2022homodyne}. The rescaled Brownian force satisfies
\begin{equation}
\langle F_{\mathrm{th}}(t)F_{\mathrm{th}}(t')\rangle
=
\bar n\,\delta(t-t'),
\label{eq:browncorr}
\end{equation}
with
\begin{equation}
\bar n=\frac{k_B T}{\hbar\Omega},
\end{equation}
which is the mean thermal phonon number of the mechanical mode in the high-temperature limit \citep{wimmer2014coherent}.

The above set of equations forms the starting point for our analysis of force transduction, added-noise spectra, and the conditions required for coherent quantum noise cancellation in the hybrid cavity magnomechanical system.
\section{Force sensing with coherent quantum noise cancellation}
\label{CQNC}

When a weak external force \(F_{\mathrm{ext}}\) acts on the mechanical resonator, it displaces the movable mirror and thereby modulates the cavity length. This modulation is imprinted onto the phase quadrature of the intracavity field and can therefore be detected through the phase quadrature of the output field. To analyze this transduction process, we solve the linearized quantum Langevin equations in the frequency domain and obtain the intracavity phase quadrature \(\hat P_a(\omega)\) (see Appendix~\ref{appendixB}). The corresponding output phase quadrature follows from the standard input-output relation,
\begin{equation}
\hat P_a^{\mathrm{out}}(\omega)=\sqrt{\kappa}\,\hat P_a(\omega)-\hat P_{a,\mathrm{in}}(\omega),
\label{eq:inputoutput_phase}
\end{equation}
where \(\kappa\) is the cavity decay rate and \(\hat P_{a,\mathrm{in}}(\omega)\) denotes the input phase quadrature.

Substituting the frequency-domain solution for \(\hat P_a(\omega)\) into Eq.~\eqref{eq:inputoutput_phase}, we obtain
\begin{align}
\label{eq:Pout_general}
&\hat P_a^{\mathrm{out}}(\omega)
=
-g\,\chi_m(\omega)\,\lambda_-(\omega)\,\sqrt{2\gamma_m\kappa}\,
\big[\hat F_{\mathrm{th}}(\omega)+F_{\mathrm{ext}}(\omega)\big]
\nonumber\\
&
+\big[g^2\chi_m(\omega)+G_{OM}^2\chi'_M(\omega)\big]
\lambda_+(\omega)\lambda_-(\omega)\,\kappa\,\hat X_{a,\mathrm{in}}(\omega)
\nonumber\\
&
+\big[\lambda_-(\omega)\kappa-1\big]\hat P_{a,\mathrm{in}}(\omega)
\nonumber\\
&
-G_{OM}\lambda_-(\omega)\sqrt{\kappa\gamma_M}
\Big[
\chi_M(\omega)\big(\Delta_M\chi'_M(\omega)+1\big)\hat X_{M,\mathrm{in}}(\omega)\\ \notag
&+\chi'_M(\omega)\hat P_{M,\mathrm{in}}(\omega)
\Big].
\end{align}

Here,
\begin{equation}
\chi_M(\omega)=\frac{1}{i\omega+\gamma_M/2},
\qquad
\chi'_M(\omega)=-\Delta_M\,\xi_M(\omega)\chi_M(\omega),
\label{eq:chiMprime_def}
\end{equation}
with \(\xi_M(\omega)\) defined in Sec.~\ref{system}. The first term in Eq.~\eqref{eq:Pout_general} contains the transduced signal and thermal force contribution, the second term represents the optical back-action contribution, the third term corresponds to the input phase noise, and the final term arises from the magnon input noise.

The back-action term can be cancelled through destructive interference between the mechanical and magnon-mediated pathways. This is the essence of coherent quantum noise cancellation (CQNC) \citep{tsang2010coherent,wimmer2014coherent,allahverdi2022homodyne,Motazedifard_2016}. The condition for complete cancellation of the back-action contribution is therefore
\begin{equation}
g^2\chi_m(\omega)+G_{OM}^2\chi'_M(\omega)=0.
\label{eq:CQNC_condition}
\end{equation}
This condition implies that the mechanical response to radiation-pressure back-action is exactly compensated by the magnon-induced contribution. Physically, it requires both a matching of susceptibilities and an appropriate balance between the effective optomechanical and cavity-magnon couplings.

Under the CQNC condition, Eq.~\eqref{eq:Pout_general} reduces to
\begin{align}
\label{eq:Pout_CQNC}
&\hat P_a^{\mathrm{out}}(\omega)
=
-g\,\chi_m(\omega)\,\lambda_-(\omega)\,\sqrt{2\gamma_m\kappa}\,
\big[\hat F_{\mathrm{th}}(\omega)+F_{\mathrm{ext}}(\omega)\big]
\nonumber\\
&\quad
+\big[\lambda_-(\omega)\kappa-1\big]\hat P_{a,\mathrm{in}}(\omega)
\nonumber\\
&\quad
-G_{OM}\lambda_-(\omega)\sqrt{\kappa\gamma_M}
\Big[
\chi_M(\omega)\big(\Delta_M\chi'_M(\omega)+1\big)\hat X_{M,\mathrm{in}}(\omega)\\ \notag
&+\chi'_M(\omega)\hat P_{M,\mathrm{in}}(\omega)
\Big].
\end{align}

It is then convenient to rewrite the output signal in terms of an equivalent added-force operator. Rearranging Eq.~\eqref{eq:Pout_CQNC}, we obtain
\begin{equation}
F_{\mathrm{ext}}(\omega)+\hat F_{\mathrm{add}}(\omega)
=
-\frac{\hat P_a^{\mathrm{out}}(\omega)}
{g\,\chi_m(\omega)\,\lambda_-(\omega)\,\sqrt{2\gamma_m\kappa}},
\label{eq:Fadd_def}
\end{equation}
where the added force noise is given by
\begin{align}
\label{eq:Fadd_final}
\hat F_{\mathrm{add}}(\omega)
&=
\hat F_{\mathrm{th}}(\omega)
-\frac{\lambda_-(\omega)\kappa-1}
{g\,\chi_m(\omega)\,\lambda_-(\omega)\,\sqrt{2\gamma_m\kappa}}\,
\hat P_{a,\mathrm{in}}(\omega)
\nonumber\\
&
+\frac{G_{OM}\sqrt{\kappa\gamma_M}}
{g\,\chi_m(\omega)\,\sqrt{2\gamma_m\kappa}}
\Big[
\chi_M(\omega)\big(\Delta_M\chi'_M(\omega)+1\big)\hat X_{M,\mathrm{in}}(\omega)\\ \notag
&+\chi'_M(\omega)\hat P_{M,\mathrm{in}}(\omega)
\Big].
\end{align}

The spectral density associated with \(\hat F_{\mathrm{add}}(\omega)\) provides a direct measure of the force sensitivity of the system under CQNC conditions \citep{Motazedifard_2016}. In the following section, we evaluate this added-force spectrum and compare the sensitivity of the present hybrid cavity magnomechanical scheme with that of conventional optomechanical sensing.
\section{Results and Discussion}
\label{sec:results}

We now turn to a quantitative analysis of the force sensitivity of the proposed hybrid system through its added-force spectral density \citep{Motazedifard_2016}. As a force-sensitivity measure, we use the force-referred noise
\begin{equation}
s_F(\omega)=\sqrt{\hbar m \Omega \gamma_m\, S_F(\omega)},
\label{eq:sensitivity_def}
\end{equation}
where \(S_F(\omega)\) denotes the dimensionless force-noise spectrum referred to the input force. A lower value of \(S_F(\omega)\) therefore corresponds to better force sensitivity.

To assess the performance of the present scheme, we compare its force-noise spectrum with that of a standard optomechanical sensor. The added-force spectrum of the hybrid system is defined as \citep{wimmer2014coherent}
\begin{equation}
S_F^{\mathrm{add}}(\omega)\,\delta(\omega-\omega')
=
\frac{1}{2}
\left[
\langle \hat F_{\mathrm{add}}(\omega)\hat F_{\mathrm{add}}(-\omega)\rangle
+\mathrm{c.c.}
\right].
\label{eq:SF_def}
\end{equation}
Using Eq.~\eqref{eq:Fadd_final}, the corresponding added-noise power spectral density takes the form
\begin{align}
\label{eq:SF_added}
S_F^{\mathrm{add}}(\omega)
=&
\frac{k_B T}{\hbar \Omega}
+
\frac{1}{2}\,
\frac{1}{g^2|\chi_m(\omega)|^2(2\gamma_m\kappa)}
\left|
\frac{\lambda_-(\omega)\kappa-1}{\lambda_-(\omega)}
\right|^2\\ \notag
& +
\frac{1}{2}\,
\frac{\omega^2+\Omega^2+\gamma_M^2/4}{\Omega^2}.
\end{align}
The first term represents the Brownian thermal contribution, the second term corresponds to optical imprecision noise, and the last term is the residual contribution associated with the magnon-assisted CQNC channel.

\begin{table}[h!]
\centering
\small
\caption{Representative experimentally accessible parameters for the hybrid cavity magnomechanical platform with a magnon ancilla.}
\label{tab:params}
\setlength{\tabcolsep}{4pt}
\resizebox{\columnwidth}{!}{%
\begin{tabular}{lcc}
\toprule
\textbf{Parameter} & \textbf{Symbol} & \textbf{Typical value} \\
\midrule
Cavity decay rate & \(\kappa\) & \(2\pi\times 1~\mathrm{MHz}\) \\
Mechanical frequency & \(\Omega\) & \(2\pi\times 10~\mathrm{MHz}\) \\
Mechanical damping rate & \(\gamma_m\) & \(2\pi\times 100~\mathrm{Hz}\) \\
Magnon linewidth & \(\gamma_M\) & \(2\pi\times 200~\mathrm{Hz}\) \\
Single-photon optomechanical coupling & \(g_0\) & \(2\pi\times 10~\mathrm{Hz}\) \\
Cavity-magnon coupling & \(G_{OM}\) & \(2\pi\times 10~\mathrm{Hz}\) \\
OPA nonlinear gain & \(\mathcal G\) & \(0.3\,\kappa\) \\
OPA phase & \(\theta\) & \(0\) -- \(\pi\) \\
Laser wavelength & \(\lambda_L\) & \(1064~\mathrm{nm}\) \\
Input power & \(P_L\) & \(10^{-3}\) -- \(10^{6}~\mathrm{pW}\) \\
\bottomrule
\end{tabular}%
}
\end{table}

In the following, we focus primarily on the measurement-induced contribution and neglect the Brownian term, as is commonly done in CQNC analyses when assessing the intrinsic quantum-limited sensitivity of the scheme \citep{Motazedifard_2016,bariani2015atom}. Under this approximation, minimizing Eq.~\eqref{eq:SF_added} yields the CQNC-limited noise floor
\begin{equation}
S_{F,\mathrm{CQNC}}(\omega)
=
\frac{1}{2}\,
\frac{\omega^2+\Omega^2+\gamma_M^2/4}{\Omega^2}.
\label{eq:SF_CQNC}
\end{equation}

For comparison, the force-noise spectrum of a standard cavity optomechanical sensor is given by \citep{RevModPhys.86.1391}
\begin{equation}
S_F^{(\mathrm{std})}(\omega)
=
\frac{k_B T}{\hbar\Omega}
+
\frac{1}{2}\,
\frac{\kappa}{\gamma_m}\,
\frac{1}{g^2|\chi_m(\omega)|^2}\,
\frac{1}{4}
+
\frac{4g^2}{\kappa\gamma_m},
\label{eq:SF_std}
\end{equation}
where the three terms arise from thermal noise, imprecision noise, and radiation-pressure back-action, respectively. Neglecting thermal noise and optimizing with respect to the drive strength gives the familiar SQL limit,
\begin{equation}
S_{F,\mathrm{SQL}}(\omega)
=
\frac{1}{\gamma_m|\chi_m(\omega)|},
\label{eq:SQL_floor}
\end{equation}
which is attained for
\begin{equation}
g_{\mathrm{SQL}}
=
\frac{\sqrt{\kappa}}{2\sqrt{|\chi_m(\omega)|}}.
\label{eq:gsql}
\end{equation}

\begin{figure}[h!]
\centering
\includegraphics[height=5cm,clip]{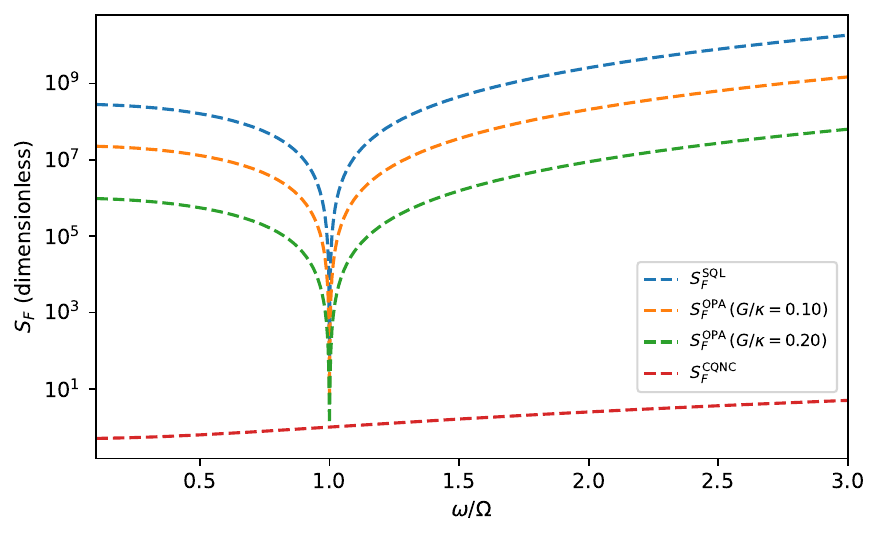}
\caption{\justifying
Force-referred noise PSD for (i) a standard optomechanical sensor (blue), (ii) an OPA-assisted hybrid sensor with \(\mathcal G/\kappa=0.1\) and \(0.3\) (orange and green), and (iii) the hybrid CQNC architecture (red). Spectra are normalized by \(\hbar m\Omega\gamma_m\), so that the plotted quantity is expressed in units of \(\mathrm{N^2\,Hz^{-1}}\). Parameters are taken from Refs.~\citep{wimmer2014coherent,singh2023enhanced}: \(g_0=300\times2\pi~\mathrm{Hz}\), \(\Omega=300\times2\pi~\mathrm{kHz}\), \(\gamma_m=30\times2\pi~\mathrm{Hz}\), \(\kappa=2\pi~\mathrm{MHz}\), \(P=100~\mathrm{mW}\), and \(\omega_L=384\times2\pi~\mathrm{THz}\).}
\label{fig:PSD1}
\end{figure}

\begin{figure}[h!]
\centering
\includegraphics[height=5cm,clip]{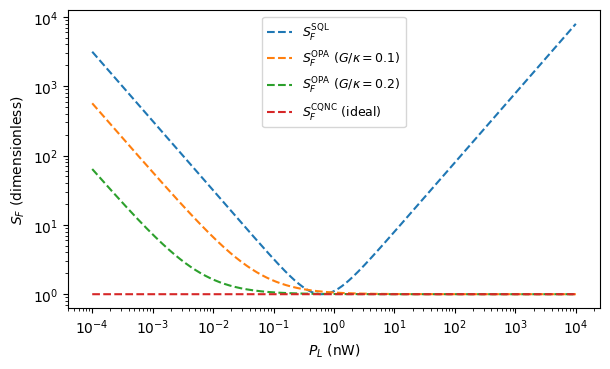}
\caption{\justifying
Resonant force-noise PSD at \(\omega=\Omega\) as a function of laser power for the standard optomechanical sensor (blue) and the OPA-assisted hybrid scheme for \(\mathcal G/\kappa=0.1\) and \(0.3\) (orange and green). The remaining parameters are the same as in Fig.~\ref{fig:PSD1}.}
\label{fig:PSD_PL}
\end{figure}

Figure~\ref{fig:PSD1} compares the spectral densities of the standard optomechanical sensor, the OPA-assisted hybrid configuration, and the CQNC-enabled hybrid system. Around the mechanical resonance \(\omega=\Omega\), all spectra exhibit a pronounced minimum. Near resonance, the OPA-assisted hybrid system already improves the sensitivity relative to the standard optomechanical case. Once the CQNC condition is imposed, the suppression of radiation-pressure back-action becomes much more pronounced, and the resulting noise spectrum lies well below the standard SQL reference over a broad off-resonant frequency range. This demonstrates that the magnon-assisted CQNC channel can significantly extend the beloew-SQL operating bandwidth.

The role of the OPA is further illustrated in Fig.~\ref{fig:PSD_PL}, which shows the resonant force-noise PSD as a function of laser power. For the standard optomechanical sensor, the noise initially decreases with increasing power because of the reduction of imprecision noise, reaches an optimum, and then rises again as radiation-pressure back-action becomes dominant. In contrast, the OPA-assisted hybrid configuration with CQNC maintains a decreasing trend up to its optimal operating point, without exhibiting the same pronounced back-action-induced upturn. Moreover, increasing the OPA gain shifts the minimum of the noise curve toward lower input power. This indicates that strong force sensitivity can be achieved in a lower-power regime, which is advantageous from an experimental perspective because it reduces the impact of heating and other power-related limitations \citep{gavartin2012hybrid}.

Taken together, these results show that the combined action of the magnon ancilla and intracavity OPA yields a substantial reduction of the force-noise spectral density compared with a conventional optomechanical sensor. The proposed hybrid scheme therefore provides a viable route toward broadband and low-power below-SQL force sensing.

\section{Imperfect Matching}
\label{sec:imperfect}

The ideal CQNC condition requires precise matching of both the relevant susceptibilities and the effective coupling strengths. In practice, however, exact matching is difficult to achieve experimentally. The dissipation rates of the mechanical and magnon modes generally cannot be tuned independently with arbitrary precision, and the effective couplings are likewise subject to finite control accuracy. It is therefore important to examine the robustness of the scheme against realistic parameter mismatch.

\subsection*{(i) Decay-rate mismatch: \(\gamma_M\neq \gamma_m\)}

We first consider the case in which all CQNC conditions are satisfied except for the matching of the decay rates. To quantify the deviation from the ideal case, we introduce the relative mismatch parameter
\begin{equation}
\delta=\frac{\gamma_M-\gamma_m}{\gamma_m}.
\label{eq:delta_def}
\end{equation}
When \(\delta\neq 0\), the magnon linewidth differs from the mechanical damping rate, and the noise floor is limited by the CQNC,
\begin{align}
S_{F,\mathrm{CQNC}}(\omega)
=&
\frac{1}{2}\,
\frac{\omega^2+\Omega^2+\gamma_M^2/4}{\Omega^2}\\
&+
\frac{\kappa g^2}{\gamma_m}\,
|\lambda_+(\omega)|^2
\left|
1+\frac{\chi'_M(\omega)}{\chi_m(\omega)}
\right|^2.
\label{eq:CQNC_mismatch_gamma}
\end{align}

\begin{figure}[!ht]
\centering
\includegraphics[width=.5\textwidth]{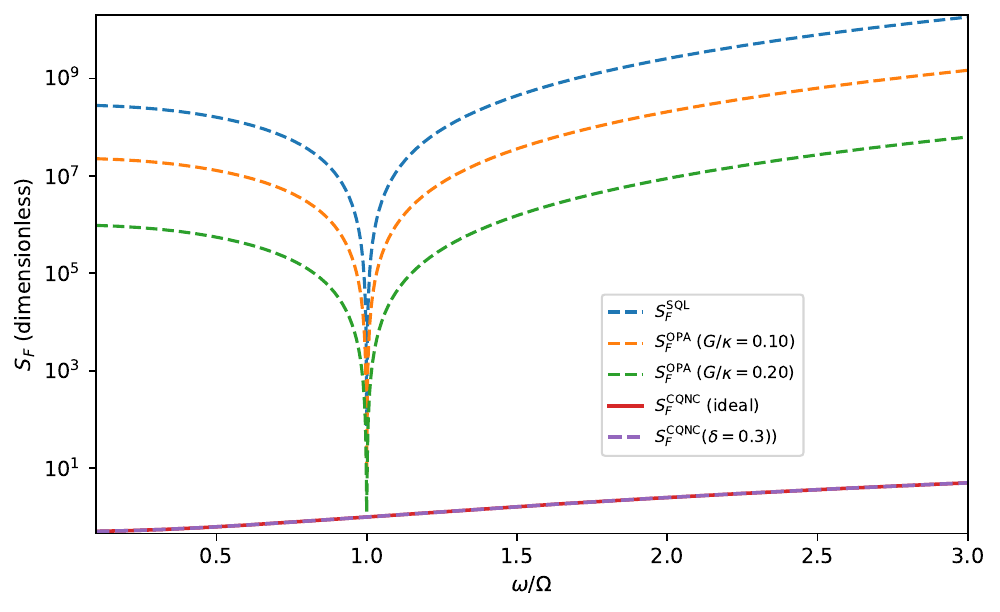}
\caption{\justifying
Variation of the force-noise spectral density with frequency in the presence of decay-rate mismatch. The blue solid line denotes the SQL, the orange and green dashed lines show the hybrid model with OPA gains \(\mathcal G/\kappa=0.1\) and \(0.2\), respectively, the red solid line corresponds to ideal CQNC, and the black dashed line represents the mismatched CQNC case with \(\delta=0.3\).}
\label{fig:Imperfect_gamma1}
\end{figure}

The corresponding spectra are shown in Fig.~\ref{fig:Imperfect_gamma1}. The ideal CQNC curve and the spectrum obtained in the presence of a moderate decay-rate mismatch remain very close to one another, indicating that the performance degradation is weak. Even for \(\delta=0.3\), the added-force noise remains near the ideal CQNC benchmark and stays substantially below the SQL over the relevant frequency range. This shows that the proposed scheme is relatively insensitive to moderate mismatch in the dissipation rates.

\subsection*{(ii) Coupling mismatch: \(G_{OM}'\neq g\)}

We next consider imperfect matching of the effective couplings. Let
\begin{equation}
\epsilon=\frac{G_{OM}'-g}{g},
\label{eq:epsilon_def}
\end{equation}
where \(G_{OM}'\) denotes the effective cavity-magnon coupling introduced in Sec.~\ref{system}. When \(\epsilon\neq 0\), the coupling-matching condition is not exactly satisfied, and the force-noise spectrum acquires a residual back-action contribution. In this case, the CQNC-limited spectrum becomes
\begin{equation}
S_{F,\mathrm{CQNC}}(\omega)
=
\frac{1}{2}\,
\frac{\omega^2+\Omega^2+\gamma_M^2/4}{\Omega^2}
+
\frac{\kappa g^2}{\gamma_m}\,
|\lambda_+(\omega)|^2
\left|
1-\frac{G_{OM}'^{\,2}}{g^2}
\right|^2.
\label{eq:CQNC_mismatch_G}
\end{equation}

\begin{figure}[!t]
\centering
\includegraphics[width=.5\textwidth]{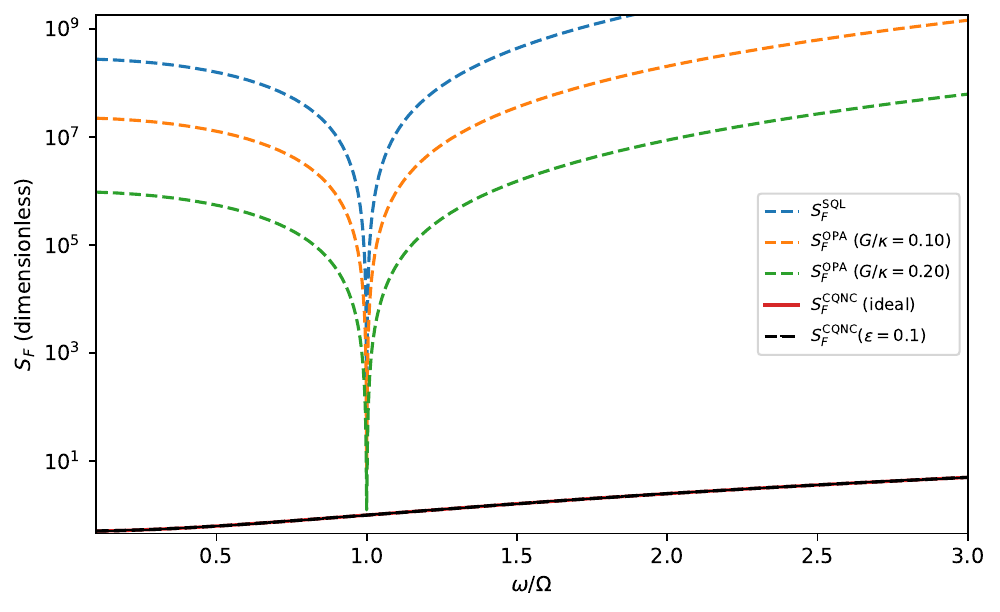}
\caption{\justifying
Variation of the force-noise spectral density with frequency in the presence of coupling mismatch. The blue solid line denotes the SQL, the orange and green dashed lines show the hybrid model with OPA gains \(\mathcal G/\kappa=0.1\) and \(0.2\), respectively, the red solid line corresponds to ideal CQNC, and the black solid line represents the mismatched CQNC case with \(\epsilon=0.01\).}
\label{fig:Imperfect_G}
\end{figure}

The spectra are plotted in Fig.~\ref{fig:Imperfect_G}. As expected, the coupling mismatch introduces a residual back-action term, so that the imperfectly matched curve lies above the ideal CQNC result. Nevertheless, even in the presence of this mismatch, the hybrid scheme continues to outperform the off-resonant SQL by several orders of magnitude. Thus, although precise coupling matching is beneficial for optimal performance, moderate deviations do not destroy the main advantage of the CQNC mechanism.

Overall, Figs.~\ref{fig:Imperfect_gamma1} and \ref{fig:Imperfect_G} show that the proposed hybrid cavity magnomechanical platform retains the principal advantages of CQNC even when the ideal matching conditions are only approximately satisfied. This robustness strengthens the practical relevance of the scheme for experimentally realistic implementations.

\section*{Conclusion}

In this work, we have investigated a hybrid cavity magnomechanical system in which the intracavity optical field is coupled simultaneously to a mechanical oscillator and a magnon mode, with the cavity further assisted by an intracavity optical parametric amplifier (OPA). Within the linearized regime, we derived the quantum Langevin equations, transformed the dynamics into the frequency domain, and obtained the corresponding output-field response relevant for force detection.

Our analysis shows that the magnon-assisted pathway can be used to realize coherent quantum noise cancellation (CQNC), whereby the magnon-induced response destructively interferes with the radiation-pressure back-action of the mechanical mode. As a result, the added-force noise can be significantly suppressed, enabling below standard-quantum-limit (sub-SQL) force sensitivity over a broad frequency range. We further found that the presence of the intracavity OPA improves the sensing performance by reducing the imprecision background and shifting the optimal operating point toward lower input laser power. This low-power operation is particularly attractive from an experimental perspective as it helps mitigate power-related limitations while preserving enhanced sensitivity.

We also examined the robustness of the scheme against imperfect parameter matching. In particular, we showed that moderate mismatch in effective coupling strengths and dissipation rates does not destroy the main advantage of the CQNC mechanism. Even away from the ideal matching condition, the hybrid system retains substantial back-action suppression and continues to operate below the standard quantum limit over an experimentally relevant bandwidth.

In general, our results identify the OPA-assisted cavity magnomechanical platform as a promising setting for back-action-suppressed weak-force detection. More broadly, the present work highlights how magnonic degrees of freedom and phase-sensitive optical control can be combined within a hybrid architecture to enhance precision sensing. These features make the proposed system relevant for applications in quantum-limited metrology, weak-force and acceleration sensing, and related hybrid quantum transduction platforms.

\section*{Acknowledgement}
A.R. gratefully acknowledges the support of a research fellowship
from UGC, Government of India. A.K.S. acknowledges the
grant from MoE, Government of India (Grant No. MoESTARS/STARS-2/2023-0161).

\section{Appendix: Derivation of the cavity phase quadrature \texorpdfstring{$\hat P_a(\omega)$}{Pa(omega)}}
\label{appendixB}

In this appendix, we derive the frequency-domain expression for the intracavity phase quadrature \(\hat P_a(\omega)\), which is used in the main text to obtain the output phase quadrature and the corresponding added-force noise spectrum.

Starting from the linearized quantum Langevin equations in the frequency domain, the relevant quadrature equations are
\begin{align}
&\hat X_b(\omega)
=
\chi_m(\omega)
\left[
-g\,\hat X_a(\omega)
+
\sqrt{2\gamma_m}\,\hat F(\omega)
\right],
\label{eq:app_Xb}
\\
&\hat X_a(\omega)
=
\sqrt{\kappa}\,\lambda_+(\omega)\,\hat X_{a,\mathrm{in}}(\omega),
\label{eq:app_Xa}
\\
&\hat P_a(\omega)
=
-g\,\chi_m(\omega)\lambda_-(\omega)
\sqrt{2\gamma_m}\,\hat F(\omega)
+
g^2\chi_m(\omega)\lambda_-(\omega)\\
&*\lambda_+(\omega)\sqrt{\kappa}\,\hat X_{a,\mathrm{in}}(\omega)
-G_{OM}\lambda_-(\omega)\hat X_M(\omega)
+
\lambda_-(\omega)\sqrt{\kappa}\,\hat P_{a,\mathrm{in}}(\omega),
\label{eq:app_Pa_start}
\\
&\hat X_M(\omega)
=
-\Delta_M\chi_M(\omega)\hat P_M(\omega)
+
\chi_M(\omega)\sqrt{\gamma_M}\,\hat X_{M,\mathrm{in}}(\omega),
\label{eq:app_XM}
\\
&\hat P_M(\omega)
=
-G_{OM}\xi_M(\omega)\lambda_+(\omega)\sqrt{\kappa}\,\hat X_{a,\mathrm{in}}(\omega)
-\chi'_M(\omega)\sqrt{\gamma_M}\,\hat X_{M,\mathrm{in}}(\omega)\\
&+
\xi_M(\omega)\sqrt{\gamma_M}\,\hat P_{M,\mathrm{in}}(\omega),
\label{eq:app_PM}
\end{align}
where
\begin{equation}
\hat F(\omega)=\hat F_{\mathrm{th}}(\omega)+F_{\mathrm{ext}}(\omega).
\end{equation}

The susceptibilities and auxiliary response functions are defined as
\begin{align}
\chi_m(\omega)
&=
\frac{\Omega}{\Omega^2-\omega^2+i\gamma_m\omega},
\\
\chi_M(\omega)
&=
\frac{1}{i\omega+\gamma_M/2},
\\
\xi_M(\omega)
&=
\left(
i\omega+\frac{\gamma_M}{2}+\Delta_M^2\chi_M(\omega)
\right)^{-1},
\\
\chi'_M(\omega)
&=
-\Delta_M\,\xi_M(\omega)\chi_M(\omega),
\\
\lambda_\pm(\omega)
&=
\left(\chi_a^{-1}(\omega)\mp 2\mathcal G\right)^{-1}.
\end{align}

To obtain a closed expression for \(\hat P_a(\omega)\), we first substitute Eq.~\eqref{eq:app_PM} into Eq.~\eqref{eq:app_XM}. This gives
\begin{align}
\hat X_M(\omega)
&=
-\Delta_M\chi_M(\omega)
\left[
\begin{aligned}
&-G_{OM}\xi_M(\omega)\lambda_+(\omega)\sqrt{\kappa}\,\hat X_{a,\mathrm{in}}(\omega)\\
&-\chi'_M(\omega)\sqrt{\gamma_M}\,\hat X_{M,\mathrm{in}}(\omega)\\
&+\xi_M(\omega)\sqrt{\gamma_M}\,\hat P_{M,\mathrm{in}}(\omega)
\end{aligned}
\right]
\nonumber\\
&\quad
+
\chi_M(\omega)\sqrt{\gamma_M}\,\hat X_{M,\mathrm{in}}(\omega).
\end{align}

Using the definition
\[
\chi'_M(\omega)=-\Delta_M\xi_M(\omega)\chi_M(\omega),
\]
this simplifies to
\begin{align}
\hat X_M(\omega)
&=
G_{OM}\chi'_M(\omega)\lambda_+(\omega)\sqrt{\kappa}\,\hat X_{a,\mathrm{in}}(\omega)\notag\\&
+
\chi_M(\omega)\bigl[1+\Delta_M\chi'_M(\omega)\bigr]\sqrt{\gamma_M}\,\hat X_{M,\mathrm{in}}(\omega)
\nonumber\\
&\quad
-\chi'_M(\omega)\sqrt{\gamma_M}\,\hat P_{M,\mathrm{in}}(\omega).
\label{eq:app_XM_closed}
\end{align}

Substituting Eq.~\eqref{eq:app_XM_closed} into Eq.~\eqref{eq:app_Pa_start}, we obtain
\begin{align}
\hat P_a(\omega)
&=
g^2\chi_m(\omega)\lambda_-(\omega)\lambda_+(\omega)\sqrt{\kappa}\,\hat X_{a,\mathrm{in}}(\omega)\\&\notag
-g\,\chi_m(\omega)\lambda_-(\omega)\sqrt{2\gamma_m}\,\hat F(\omega)
\nonumber\\
&\quad
-G_{OM}\lambda_-(\omega)
\Big[
G_{OM}\chi'_M(\omega)\lambda_+(\omega)\sqrt{\kappa}\,\hat X_{a,\mathrm{in}}(\omega)\notag \\&
\quad+
\chi_M(\omega)\bigl(1+\Delta_M\chi'_M(\omega)\bigr)\sqrt{\gamma_M}\,\hat X_{M,\mathrm{in}}(\omega)
\nonumber \\
&\quad -\chi'_M(\omega)\sqrt{\gamma_M}\,\hat P_{M,\mathrm{in}}(\omega)
\Big]
+
\lambda_-(\omega)\sqrt{\kappa}\,\hat P_{a,\mathrm{in}}(\omega).
\end{align}
Collecting terms finally yields
\begin{equation}
\begin{aligned}
\hat P_a(\omega)
&=
\Big[
 g^2\chi_m(\omega)+G_{OM}^2\chi'_M(\omega)
\Big]
\lambda_-(\omega)\lambda_+(\omega)\sqrt{\kappa}\,\hat X_{a,\mathrm{in}}(\omega)\\
&\quad
-g\,\chi_m(\omega)\lambda_-(\omega)\sqrt{2\gamma_m}\,\hat F(\omega)
+\lambda_-(\omega)\sqrt{\kappa}\,\hat P_{a,\mathrm{in}}(\omega)\\
&\quad
-G_{OM}\lambda_-(\omega)\sqrt{\gamma_M}
\Big[
\chi_M(\omega)\bigl(1+\Delta_M\chi'_M(\omega)\bigr)\hat X_{M,\mathrm{in}}(\omega)
-\chi'_M(\omega)\hat P_{M,\mathrm{in}}(\omega)
\Big].
\end{aligned}
\label{eq:Pa_omega_appendix}
\end{equation}

Equation~\eqref{eq:Pa_omega_appendix} is the required expression for the intracavity phase quadrature in terms of the input noise operators and the total force. Using the standard input-output relation,
\begin{equation}
\hat P_a^{\mathrm{out}}(\omega)=\sqrt{\kappa}\,\hat P_a(\omega)-\hat P_{a,\mathrm{in}}(\omega),
\end{equation}
one then obtains the output phase quadrature used in the main text to derive the added-force noise spectrum and the corresponding force-sensing sensitivity.
\bibliography{bibb}
\end{document}